\documentclass[final,3p,times,twocolumn]{elsarticle}
 \biboptions{comma,sort&compress}
\usepackage{here}
 \usepackage{graphicx}
  \usepackage{epsfig}

\def\nin{\noindent}
\def\beq{\begin{equation}}
\def\eeq{\end{equation}}
\def\bea{\begin{eqnarray}}
\def\eea{\end{eqnarray}}

\journal{Nuc. Phys. (Proc. Suppl.)}

\begin{document}

\begin{frontmatter}



\title{ High orders perturbation theory and dual models for Yang-Mills theories}

 \author[label1]{Valentin Zakharov\corref{cor1}}
  \address[label1]{ Institute of Theoretical and Experimental Physics
\\
 Bolshaya Cheremushkinskaya 25, Moscow, Russia}
\cortext[cor1]{Speaker}
\ead{vzakharov@itep.ru}



\begin{abstract}
\noindent
 We start with the QCD sum rules  which are originally based on the idea
 that it is the power-like    corrections to the parton model
 which are related to the confinement. The naive use of the Operator Product Expansion
 ensures that there is  a 'gap'
  in the powers of $\Lambda_{QCD}$   which miss the quadratic terms and start with
  the  quartic term, proportional to the gluon condensate, $<(G_{\mu\nu}^a)^2>$.
  We review how this hypothesis
 stood against various checks through the last three decades and
 how it was modified  through inclusion
 of the missing link, that is quadratic corrections.
 In field theoretic language the quadratic corrections are
 dual to long perturbative series.
 In the dual description, the quadratic
 corrections are conveniently parameterized  in terms of the metric in extra dimensions.
  We emphasize that  the dual models do {\it not} incorporate the
 so called infrared renormalon.
\end{abstract}


\end{frontmatter}


\section{Introduction}
\nin
This talk was given at a special session
of the QCD10 conference at Montpellier,
devoted to (30+1) year of the QCD sum rules \cite{svz}.
I am thankful to Stephan Narison for organizing this event
and, also, for overtaking the most difficult part
of the job, that is reviewing the
present status  of the sum rules \cite{snb}.
 As for myself,   I chose for myself
to talk more about not yet fully settled
issues.
The emphasize is mainly on the quadratic
corrections, simply missing from the standard sum rules.
To a large extent the talk is based
on the published papers \cite{narison,andreev4}
but we add a few remarks as well.

 Originally the sum rules were applied to two-point
 functions induced by external currents.
 More specifically, one considers integrals of the kind
 \begin{equation}\label{observable}
 f(M^2)~\equiv~\int ds \exp(-s/M^2)Im \Pi(s)~,
 \end{equation} here $M^2$ is a  large mass parameter,  $Im \Pi(s)$
 is imaginary part of a polarization operator
 and, speaking generically, can be
 measured by studying transition induced by currents.
 The sum rules equate the
 observable (\ref{observable}) to power corrections:
 \begin{equation}\label{symbolic}
 f(M^2)~\approx~[f(M^2)]_{parton~model}\big(1+c_{gl.cond}{\Lambda_{QCD}^4\over M^4}\big)
 \end{equation}
 where the product $c_{gl.cond}\Lambda_{QCD}^4$ is  calculable
 in terms of the gluon condensate,
 or matrix element $<(G_{\mu\nu}^a)^2>$.

 The form (\ref{symbolic}) is oversimplified since it omits,
 in particular, the first perturbative
 correction of order $\alpha_s(M^2)\sim\ln^{-1}M^2$, the
 quark-condensate terms and so on.
 But nevertheless, eq. (\ref{symbolic}) does
 summarize correctly the basic idea that
 in the crucial region of $M^2$ it is the
 power corrections which are related to the
 confinement (manifested through resonance masses
 and widths entering (\ref{observable})).
 The power corrections are given in terms of matrix elements of
 local gauge invariant operators.
 In particular, there is no correction of order
 $\Lambda_{QCD}^2$ since there is no
 corresponding operator of dimension d=2. Eq.
 (\ref{symbolic}) is in no way obvious
 and was introduced on phenomenological grounds \cite{svz}.
 In particular, dropping the perturbative
 corrections which are powers of $\alpha_s(M^2)$
 and keeping the power correction which is of order
 $\exp(-const/\alpha_s)$ might look  confusing.
\section{ Quadratic corrections: \\ unifying continuum- and lattice-languages}

 The expansion (\ref{symbolic}) works well in many cases \cite{snb}.
 However, there exist reasons to  revisit it:
 \begin{enumerate}
 \item{Trying to improve the quality of the sum rules one starts to calculate
 perturbative corrections. The question
  is,
 whether  the numerical value of $<G^2>$ is kept
 independent of these corrections or should it   vary
  from  order to order of perturbation theory?}
 \item{The Cornell potential for heavy quark interaction:
 \begin{equation}
 V_{Q\bar{Q}}(R)~\approx~-{const\over R}+\sigma R~,
 \end{equation}
 where $R$ is the distance between the quarks and $\sigma$ is the string tension
 holds numerically at all the distances measured on the lattice
 including small distances.  In fact, this is an ideal example of a kind of expansion (\ref{symbolic}),
 with a single power-like correction present. The problem is that the correction
 to the leading, Coulomb-like term at short distances
 is quadratic, $\sim (\sigma R^2)$. And quadratic corrections are not
 originally included into (\ref{symbolic}) and things look mysterious.}
 \end{enumerate}

Although the questions are straightforward,
convincing answers are difficult
to get phenomenologically. The reason is that, after all,
we talk about relatively small power-like
corrections. Each particular case of
such phenomenology is tedious, and more important,
very difficult to follow from outside.
Nevertheless, after a few years of
hard work the answer seems to be unique:
quadratic corrections do exist \cite{chetyrkin}.

Moreover, the quadratic correction is directly related to confinement.
Although the statement might look too strong and vague, actually,  it has
a well defined content. Namely, on the lattice one is able to
clarify what kind of field configurations are responsible for the confinement.
In the lattice nomenclature these configurations are called monopoles and
vortices,
for a review see, e.g., \cite{greensite}.
Most amusing,  they occupy a small fraction of the
lattice which tends to zero with the vanishing lattice spacing $a\to 0$.
Closer to our story,
it was demonstrated that the non-perturbative, i.e. monopole- or vortex- related
potential is indeed linear at all the distances beginning
with a single lattice spacing:
\begin{equation}\label{potential}
\Big(V_{Q\bar{Q}}(R)\Big)_{non-perturbative}~=~\sigma ~\cdot R~.
\end{equation}
Details and references can be found in \cite{viz}.
Thus, the short-distance quadratic correction is certainly there
!
\section{Power corrections vs perturbative series}

What is the relation between the power corrections and perturbative series?
Concentrate on the gluon condensate itself. Then perturbatively:
\begin{equation}\label{pert}
a^4{\pi^2\over 12N_c}\Big({-b_0g^3\over  \beta(g)}\Big)<{\alpha_s\over \pi}GG>_{pert}=\Sigma_{n=1}^{\infty}
\big(1+a_n\alpha_s^n\big)~,
\end{equation}
where $\alpha_s\equiv g^2_s/4\pi$ is the strong interaction coupling, $\beta (g)$ is the beta function, $a_n$ are perturbative coefficients. Moreover, $a$ is the lattice spacing and
the factor $a^4$ in the l.h.s. of Eq (\ref{pert}) is
introduced to cancel the UV divergence inherent
to the quantity considered.

The perturbative expansion (\ref{pert}) is expected to be asymptotic.
Namely, for $n$ large enough the expansion coefficients are expected to
grow factorially:
\begin{equation}\label{ir}
\lim_{n\to\infty}{a_n}_{IR}~=~n!\big({b_0\over 2}\big)^n
\end{equation}
 The series (\ref{ir}) is called infrared renormalon. If we estimate  the
  uncertainty of the asymptotic expansion due to the factorial growth (\ref{ir})
 we find contribution of order
 $(\Lambda_{QCD}\cdot a)^4$ compared to the leading perturbative
 term. Such an uncertainty would introduce
 $<G^2>\sim \Lambda_{QCD}^4$ which is
 the physical gluon condensate entering,
 in particular QCD sum rules (\ref{symbolic}).
 This is the standard wisdom on the relation between
 divergences of the perturbative
 series in large orders $n$ and the Operator Product Expansion.

 Where is then the hypothetical quadratic correction? Well, there is no
 pronounced role for such a correction
 in the set up considered. It should be buried within the still-convergent
 orders of the perturbative series.
 In other words, keeping the quadratic correction might be reasonable only as far
 as the perturbative series is not long enough.
 If we keep many terms, then the
 quadratic correction   is to be eaten up by the perturbative terms.
  Note that within such a logic the quadratic correction (if any) is inferior
  to many orders of perturbation series and uninteresting. Moreover, the very language
  of power corrections seems rather irrelevant
  if we need to keep explicit many orders of perturbation theory.

  Things become, however, much more interesting if we turn to the example of the
  longest perturbative series known.
  It is indeed for the gluon condensate (\ref{pert})
  and contains 20 (no mistake: twenty) first terms in the expansion,
  see \cite{rakow} and references therein.
  Moreover, the full value of the gluon
  condensate is known from the lattice measurements
  since it is simply the plaquette action and
  can be measured to a very high precision.
  As a result one can use the following fitting procedure:
  \begin{equation}
  (\Delta P)_N\equiv P_{full}-\Sigma_n p_n\alpha_s^n~\approx~(\Lambda_{QCD}\cdot a)^{\rho(N)}~,
  \end{equation}
  where $P_{full}$ is the exact (or full) plaquette action,
  $p_n$ are perturbative coefficients
  evaluated explicitly and the difference
  between the partial (up to order N) perturbative
 series  and the full value is fitted by a
 power-like correction where the  index
 $\rho(N)$ depends   on $N$ itself.

 The results \cite{rakow} concerning the
index $\rho(N)$  are remarkable.
Namely, the fits produce:
 \begin{eqnarray}
 \rho(N)~\approx~2~~if~N~\leq~10\\ \nonumber
 \rho(N)~\approx~4~~if~N~\geq~10\nonumber
 \end{eqnarray}
 Thus the perturbative series does know about
 the sacred quadratic and quartic corrections
 although this knowledge is very difficult
 to express analytically. Everything is numerical.

 Another remarkable result is the simplicity of the coefficients $p_n$.
 The expansion is close to
a geometric series with the ratio
\begin{equation}\label{ratio}
r_n~\equiv~{p_{n+1}\over p_n} ~\approx~Const~~.
\end{equation}
 and we do not quote the numericalvalues of the fit parameters here.
  Observation (\ref{ratio}) implies that
  {\it there is no sign of the infrared renormalon
  at least in the first 20 orders of expansion}.
  This is the reality which we have to confront and appreciate theoretically.

  In conclusion of this section it is worth mentioning that similar interplay between
  perturbative and power-like corrections was observed earlier by analyzing much
  shorter perturbative expansions. For example, it was observed in Ref. \cite{necco}
  that the quadratic correction at short distances to the Coulomb-like potential
  (see Eq. (\ref{potential})) is reproduced in fact by higher orders of perturbation
  theory. Another example
  is provided by Ref. \cite{kataev} where   sum rules for the structure function $xF_3(x)$
   measurable in deep inelastic neutrino scattering were analyzed. It turned out
   that including higher orders of the perturbative expansion reduces  considerably
   the numerical value of the quadratic correction. Note that in case of the
   $xF_3(x)$ function the quadratic correction is commonly associated with the
   infrared renormalon while in case of the quark potential the quadratic correction
   corresponds to the ultraviolet renormalon \footnote{Mostly, when we talk about
   the quadratic corrections we have in mind vacuum-to-vacuum matrix element
   of a correlator of two
   currents. Then the quadratic correction might come only from short distances and has
   no straightforward interpretation in terms of the OPE. In case of the deep inelastic scattering,
   as is well known, quadratic correction is associated with large distances.}. Strong correlation between the quadratic
   correction and perturbative series is found in both cases. The limitation of the
   analyses just mentioned is that the perturbative series known explicitly are relatively
   short, say 3-4 terms, and to reach conclusions one relies on theoretical estimates of higher-order terms. Signals for importance of quadratic corrections were
   obtained in quite a few other analyses, see in particular \cite{megias}.

   To summarize, there is strong evidence in favor of crucial role of the
   unconventional quadratic corrections
   in correlators of two currents. Moreover, the quadratic correction is dual to a long
   perturbative series. Thus, one is to use either of them but not both. "Unconventional"
   means that there is no operator of the dimension $d=2$ which would enter the OPE. Thus, it
   is difficult to even parameterize the quadratic correction within the field-theoretic approach:
   the quadratic correction is a part of the coefficient function in front of the unit operator
   but we are lacking means to distinguish it from the rest of the coefficient function.

   \section{Dual models and power-like corrections}
   \subsection{Finding a dimension-two parameter}
The situation with the quadratic correction is quite paradoxical.
On one hand, there is accumulating and strong phenomenological evidence in favor
of such a correction. On the other hand, the field-theoretic framework
does not even provide us with a suitable parameterization of the quadratic correction.
The most successful phenomenological model \cite{chetyrkin}
introduces a "short-distance
gluon mass" $m_g$, so that one replaces the gluon propagator by
\begin{equation} \label{replacement}
{1\over q^2}~\to~{1\over q^2+m_g^2}~,
\end{equation}
where the gluon mass turns to be tachyonic.
Clearly, the replacement (\ref{replacement})
is consistent with the gauge invariance only in the Born approximation.
In higher orders, there is no way to introduce a dimension-two quantity.

The resolution of the paradox seems to be that the quadratic correction belongs rather to the
realm of the dual models for the Yang-Mills theories.
As is well known, the dual or stringy formulations of non-Abelian gauge theories
is derived only in some supersymmetric cases. In QCD case, there are educated
guesses on the universality class \cite{witten,sugimoto} and more explicit but rather
arbitrary models, see, in particular, \cite{karch,andreev,andreev1}.
The main point now is that the  models allow to
introduce a dimension-two parameter to the theory in a gauge invariant and
natural way.

To begin with, the stringy models readily incorporate confinement.
Indeed, linear potential at large distances, see (\ref{potential}),
is commonly related to existence of a string connecting the quarks.
To be more ambitious and describe the potential at all the distances
one can introduce  a running string tension
$\sigma(l)$ where $l$ is the length of the string and $\sigma(l)$ is its tension
as function of the length. Moreover, one can trade the running tension
for a string living in  an extra dimension $z$ with non-trivial geometry, $G_{mn}(z)$.
The action of the string is the Goto-Nambu action,
 \begin{equation}\label{nambu}
 S~=~{1\over 2\pi \alpha^{'}}\int d^2\xi \sqrt{det G_{nm}\partial_{\alpha}X^m\partial_{\beta}X^n}~.
 \end{equation}
 The relation to the potential is provided by the boundary condition
 that the string ends on the Wilson line \cite{rey},
 \begin{equation}\label{wilson}
 <W(C)>~=~exp(-S(C)_{min}),
 \end{equation}
 and the Wilson line belongs to  $z=0$.
 One can then trade the potential (\ref{potential}) for an explicit form of
 the metric $G_{mn}(z)$.
 In particular, the following metric reproduces
 the Cornell potential (\ref{potential})
 quite satisfactory \cite{andreev1}:
\begin{equation}\label{metric}
ds^2\equiv G_{mn}dX^mdX^n=R^2{e^{cz^2/2}\over z^2}(dx^idx^i+dz^2),
\end{equation}
where the ordinary 4 dimensional space corresponds to $z=0$, $R^2$ is a constant,
the parameter $c$ is of phenomenological nature and its numerical value,
$c\approx 0.9 ~GeV^2$, is
obtained from the fitting procedure.

Most remarkably, the parameter $c$ specifies the quadratic correction
at small $z$ in a perfectly
gauge invariant way. Thus, the metric (\ref{metric}) allows to introduce
the quadratic correction in the universal geometric language.
Upon fixation  of the coefficient $c$ in terms of the potential one
can, in principle, evaluate quadratic corrections to other observables.
For technical reasons, there are not many examples of this kind.
However, one  can say that
the emerging phenomenology turns to be successful, see, in particular
\cite{andreev,andreev2}.

To summarize, the use of the dual models  promotes the numerical observation
on the dominance of the quadratic correction at small $R$
in the heavy-quark potential (\ref{potential}) to the status of a universal feature
of the short-distance physics.

\subsection{ On the sign of the quadratic correction}

There is a puzzling question about the sign of the coefficient $c$ in the metric (\ref{metric}).
We fixed it as positive, by mapping the potential (\ref{potential}) into the metric.
Actually, the sign is fixed  merely by the condition that we should have confinement,
as first observed in \cite{andreev1}, see also
\cite{brodsky}. On the other hand, papers which introduced the quadratic
correction originally \cite{karch,andreev} are using rather negative value
of the coefficient
$c$, $h\sim exp(-cz^2/2)$, instead of $h\sim exp(+cz^2/2)$ as in Eq. (\ref{metric}).

This discrepance might be
finally settled in favor of, say, positive sign of $c$ through
further phenomenological analysis. One cannot rule out, however, that there is a matter
of principle behind this apparent discrepancy.
Namely, the papers \cite{karch,andreev} refer rather to Minkowski space
while numerically impressive successful applications \cite{andreev1,andreev2}
of the metric
 refer to the Euclidean space.
Then we would come to an intriguing possibility that in theories with confinement
there is no analytical continuation of the quadratic correction
from the Minkowski to Euclidean space.

The question is then, whether the insight we got on the nature of the quadratic correction as being
dual to long perturbative series supports the idea of the non-analyticity or rather rejects it.
Analytic properties of the running coupling are discussed in many papers, see,
in particular, \cite{dokshitzer} and it seems obvious that the continuation of the coupling from
the Minkowski to Euclidean space and vice verse is not straightforward at all.

To put it even simpler, by continuing the log factor in the running coupling
one gets an imaginary part in the Minkowski region, $\ln Q^2=\ln(-Q^2)+i\pi$.
The imaginary part is big numerically and summing up higher orders in perturbation
series at $Q^2>0$ and at $q^2=-Q^2$ might well result in different
 quadratic corrections.

Basing on  the example of the perturbative evaluation of the gluon condensate
discussed above it is tempting to assume that geometric-series-type of expansion
does incorporate the quadratic correction in
other cases as well. And, indeed, one observes \cite{narison} that
existing examples of relatively long explicit perturbative expansions do not
rule out that geometric series approximately apply in other cases as well.
For example, using the results of Ref. \cite{baikov}
one finds for the polarization operator associated with
the scalar currents in QCD:
\begin{eqnarray}\label{expansion}
-Q^2{d\over dQ^2}\Pi_5(Q^2)~=~{N_c\over 8\pi^2}\Big(1+5.67\alpha_s\\ \nonumber
+45.85\alpha_s^2+465.8\alpha_s^3+5589\alpha_s^4\Big)~.
\end{eqnarray}
We see that this expansion numerically is quite close to a geometric series.
In the context of our discussion, it is crucial that the expansion (\ref{expansion})
refers to the Euclidean coupling constant $\alpha_s(Q^2)$. If one would use
$\alpha_s(-Q^2)$ as an expansion parameter, no similarity to a geometric series
would persist.

To summarize, the duality between the power corrections and long perturbative series
rather supports the suggestion that the quadratic correction cannot be trivially
continued from the Minkowskian to Euclidean domain and vice verse. Much more is
to be done, however, to strengthen the argumentation.

\subsection{Absence of infrared renormalon}

Probably one of the most important changes brought by the dual models
to the dogma of high-order corrections is the {\it absence of the infrared renormalon}
\cite{andreev4}.

In more detail, to imitate the gluon condensate (\ref{pert}) in the continuum theory
one  evaluates vacuum expectation value of a circular Wilson line of
small radius $r$, $r^2\Lambda_{QCD}^2\ll 1$.
In the dual model one uses eqs (\ref{wilson}), (\ref{nambu}), (\ref{metric})
to perform the explicit calculation. The result can be represented as:
\begin{equation}\label{andreeev}
<W(r)>\approx\exp\Big[const\big(1-0.84 cr^2-0.035 c^2r^4\big)\Big].
\end{equation}
The first two terms in the r.h.s. of (\ref{andreeev}) correspond to
the quartic
and quadratic ultraviolet divergences  in (\ref{pert}). The evaluation
of these terms in the continuum theory  is not unique and depends in fact on the regularization
procedure. These terms cannot be compared directly with the results of the
perturbative calculation (\ref{pert}) in the lattice regularization.

The most interesting result is the term proportional to $c^2$ in the r.h.s.
of Eq (\ref{andreeev}).
It corresponds, in the continuum language to $<G^2>\sim \Lambda_{QCD}^4$
and is most relevant to the sum rules (\ref{symbolic}) and phenomenological
fits. Moreover, it is this term which is commonly related to the
expected IR-renormalon divergence of the perturbative series (\ref{pert}).
The crucial observation \cite{andreev4} is that in the dual-model approach
the $<G^2>\sim \Lambda_{QCD}^4$ term is related in fact to short distances
and is calculable. It corresponds to the $z^4$ term in the expansion
of the factor $h(z)=\exp(cz^2/2)$ in the metric (\ref{metric}) at small
$z$, $z\to 0$.

This is quite a remarkable possibility revealed by the dual models:
the $<G^2>\sim \Lambda_{QCD}^4$ piece in the gluon condensate is
associated with the small, not large distances.
If the piece $<G^2>\sim \Lambda_{QCD}^4$ is indeed associated with
short distances then it should be calculable in perturbation theory.
This prediction of the
dual models is in perfect agreement with the absence of the infrared renormalon
from the explicit calculation (\ref{pert}), see for discussion above.
The  ifference between the standard field theoretic estimates
which indicate the presence of the IR renormalon and dual approach which,
 does not incorporate the infrared-sensitive piece
of $<G^2>$ is of pure geometric nature. In the both approaches,
the total value of the small Wilson line is dominated by small
distances, of order $r$. The infrared sensitive piece is
anyhow relatively small and is due to rather
exceptional or suppressed configurations which involve  distances
much larger than $r$. One can readily check that it is much easier to reach
such distances for one virtual particle (gluon) than for a string,
see (\ref{wilson}), (\ref{metric}).

\subsection{Choice of the dual model}

We can summarize our discussion by saying that the mystery of the quadratic
correction is resolved only by the dual models. Moreover, one can reverse the logic
and try to clarify which dual models are fitting the knowledge on the power
corrections in the best way. The conclusions then are:
\begin{enumerate}
\item{the quadratic correction is to be built   explicitly into   the dual
model}
\item{the model is to be formulated in terms of strings, not just fields living
in the extra dimensions}
\end{enumerate}
It is interesting to note that an independent and much more thorough analysis
of phenomenological implications of the
existing dual models results in a similar conclusion \cite{reece}.
Unfortunately, stringy models are more difficult to apply to
phenomenology of the power corrections and mostly one considers
models with fields (not strings) living in extra dimensions,
see, e.g., \cite{nikotri}.

\section{Conclusions}
\nin
To summarize, there is substantial progress in understanding of the power
corrections brought to light by the sum rules. Namely, the emphasize shifted
to the quadratic correction  absent from the original, simplified form of the
sum rules. It turned out that it is just this correction which is most closely
related to the confinement.  Moreover, introduction of this correction
allows for a straightforward interpretation of some lattice data and unifies
the continuum-theory and lattice languages. Thus, phenomenologically the
quadratic correction resolves quite a few puzzles. However, interpretation
in the field-theoretic language is rather awkward: it is dual to a long
perturbative series, as is confirmed by a number of perturbative calculations,
see in particular
\cite{necco,kataev,rakow}. With the advent of the dual models of QCD
the quadratic correction found its interpretation in terms of the metric
in an extra coordinate $z$. Dual models which incorporate this quadratic correction
at small $z$ turn to be successful phenomenologically, probably even most successful.
\section*{Acknowledgements}
\nin
 The author would like to thank  S. Narison for the invitation and hospitality.
 This work was partially supported by RFBR grant no. 10-02-01483.


\begin{thebibliography}{999}
\vspace*{-0.25cm}
\bibitem{svz} M.A. Shifman, A.I. Vainshtein and V.I. Zakharov,
Nucl. Phys. {\bf B147} (1979) 385.

\bibitem{snb} For a thorough review see S.
Narison, {\it QCD as a theory of hadrons,
Cambridge Monogr. Part. Phys. Nucl. Phys. Cosmol.} {\bf 17}
(2002) 1-778
[hep-h/0205006].

\bibitem{narison}
S. Narison and  V.I. Zakharov,  Phys. Lett. {\bf B679}  (2009) 355,
arXiv:0906.4312 [hep-ph].

\bibitem{andreev4}
O. Andreev and V.I. Zakharov,
Phys. Rev. {\bf D76} (2007) 047705,  [arXiv:hep-ph/0703010].

\bibitem{chetyrkin}
K.G. Chetyrkin,  S. Narison and V.I. Zakharov,  Nucl. Phys. {\bf B550} (1999) 353,
[arXiv:hep-ph/9811275];\\
M.N. Chernodub, F.V. Gubarev,   M.I. Polikarpov and  V. I. Zakharov,  Phys. Lett.
{\bf B475} (2000) 303, [arXiv:hep-ph/0003006];\\
S. Narison and  V. I. Zakharov,  Phys. Lett. {\bf B522} (2001) 266, [arXiv:hep-ph/0110141].

\bibitem{greensite}
J. Greensite,  Prog. Part. Nucl. Phys. {\bf 51} (2003) 1,
[arXiv:hep-lat/0301023].

\bibitem{viz}
V.I. Zakharov,   Nucl. Phys. Proc. Suppl. {\bf 164} (2007) 240,
[arXiv:hep-ph/0509114].

\bibitem{rakow}
P.E.L. Rakow,  PoS LAT2005 {\bf 284} (2006), [arXiv:hep-lat/0510046];\\
E.-M. Ilgenfritz,  Y. Nakamura,  H. Perlt, P.E.L. Rakow,
G. Schierholz, arXiv:0910.2795 [hep-lat];\\
M. Gockeler et al.,  : arXiv:1003.5756 [hep-lat].

\bibitem{necco}
 S. Necco, R. Sommer,   Phys. Lett. {\bf B523} (2001) 135, [arXiv:hep-ph/0109093].

\bibitem{kataev}
A.L. Kataev,  G. Parente,  A.V. Sidorov,
Nucl. Phys. Proc. Suppl. {\bf 116} (2003) 105, [arXiv:hep-ph/0211151].

\bibitem{megias}
E. Megias,  E.R. Arriola, L.L. Salcedo, Nucl. Phys. Proc. Suppl. {\bf 186} (2009) 256, arXiv:0809.2044 [hep-ph];\\
S.S. Afonin,  Phys. Lett. {\bf B678}  (2009) 477, arXiv:0902.3959 [hep-ph];\\
E. Megias,   E.R. Arriola, L.L. Salcedo,  Phys. Rev. {\bf D81} (2010) 096009,  arXiv:0912.0499 [hep-ph].
\newpage

\bibitem{witten}
E. Witten, Adv. Theor. math. Phys. {\bf 2} (1998) 505.

\bibitem{sugimoto}
T. Sakai,   Sh. Sugimoto,   Prog. Theor. Phys. {\bf 113} (2005) 843 [arXiv:hep-th/0412141].

\bibitem{karch}
A. Karch,  E. Katz,   D. T. Son,  M. A. Stephanov,  Phys. Rev. {\bf D74} (2006) 015005, [arXiv:hep-ph/0602229].

\bibitem{andreev}
O. Andreev,  Phys. Rev. {\bf D73} (2006) 107901.
[arXiv:hep-th/0603170].

\bibitem{andreev1}
O. Andreev and  V. I. Zakharov,
Phys. Rev. {\bf D74} (2006) 025023,2006,  [arXiv:hep-ph/0604204];
JHEP, {\bf 0704}  (2007) 100,
[arXiv:hep-ph/0611304].

\bibitem{brodsky}
S. J. Brodsky,  G.F. de Teramond,  A. Deur,  Phys. Rev. {\bf D8} (2010) 096010,
 arXiv:1002.3948 [hep-ph].


\bibitem{rey}
J. Maldacena, Phys. Rev. Lett. {\bf 80} (1998) 4859;\\
S.-J. Rey, J.-T. Yee, Eur. Phys. J. {\bf C22} (2001) 379.

\bibitem{andreev2}
O. Andreev,  arXiv:1008.4738 [hep-ph];\\
O. Andreev,  Phys. Rev. Lett. {\bf 102} (2009) 212001, arXiv:0903.4375 [hep-ph].


\bibitem{dokshitzer}
D.V. Shirkov, I.L. Solovtsov,  [arXiv:hep-ph/9604363];\\
Yu. L. Dokshitzer,   G. Marchesini,   B.R. Webber,  Nucl. Phys. {\bf B469} (1996) 93, [arXiv:hep-ph/9512336].

\bibitem{baikov}
P.A. Baikov, K.G. Chetyrkin, J.H. K\"uhn, arXiv:0801.1821 [hep-ph].

\bibitem{reece}
C. Csaki,   M.  Reece,   J. Terning,  {\em JHEP} {\bf 0905} 067 (2009),
 arXiv:0811.3001 [hep-ph].

\bibitem{nikotri}
H.~M. Ratsimbarison,  arXiv:1009.4637 [hep-th];\\
S. Nicotri,  arXiv:1009.4829 [hep-ph].

\end{thebibliography}
\end{document}